\documentclass[aps,prb,twocolumn,nofootinbib,floatfix,bibliography,superscriptaddress]{revtex4-2}

\usepackage{amsmath}
\usepackage{amsthm}
\usepackage{amssymb}
\usepackage{graphicx}
\usepackage{tabularx}
\usepackage{color}
\usepackage{hyperref}
\usepackage{cleveref}
\usepackage{xcolor}
\usepackage[normalem]{ulem}

\newcommand{\be}{\begin{equation}}
\newcommand{\ee}{\end{equation}}
\newcommand{\bea}{\begin{eqnarray}}
\newcommand{\eea}{\end{eqnarray}}

\def\bs#1\es{\begin{split}#1\end{split}}	\def\bal#1\eal{\begin{align}#1\end{align}}

\DeclareUnicodeCharacter{2003}{\hspace{2em}} 
\begin{document}
\title{Engineering the localization transition in a Charge-Kondo circuit 
}

\author{Zhanyu Ma}
\affiliation{Raymond and Beverly Sackler School of Physics and Astronomy, Tel Aviv University, Tel Aviv 69978, Israel}

\author{Cheolhee Han}
\affiliation{Department of Data Information and Physics, Kongju National University, Kongju 32588, Republic of Korea}	

\author{F. Pierre}
\affiliation{Universit\'e Paris-Saclay, CNRS, Centre de Nanosciences et de Nanotechnologies (C2N), 91120 Palaiseau, France}

\author{Eran Sela}
\affiliation{Raymond and Beverly Sackler School of Physics and Astronomy, Tel Aviv University, Tel Aviv 69978, Israel}

\begin{abstract}
Charge Kondo circuits consist of metallic islands connected by single-mode quantum point contacts (QPCs). The island’s charging energy makes these circuits tunable quantum simulators of various strongly interacting models. Here we propose a circuit that realizes the Kondo effect with effective Luttinger-liquid interactions, and show that it undergoes a localization transition in which the QPC transmission is fully suppressed below a critical value. Experimental signatures include a diverging charge susceptibility and an entropy step. Our findings open a path toward realizing localization transitions in more exotic settings.
\end{abstract}
\maketitle

Localization transitions (LTs) arise from coupling a quantum system to an external environment, and mark a change from coherent quantum dynamics to incoherent behavior
~\cite{RevModPhys.59.1,weiss2012quantum,PhysRevLett.49.681,PhysRevLett.49.1545,PhysRevLett.51.1506}. A range of proposals have been put forward to realize such transitions in mesoscopic settings~\cite{PhysRevLett.92.196804,PhysRevB.72.073305,PhysRevLett.95.086406,PhysRevLett.97.016802,PhysRevA.78.010101,PhysRevLett.104.226805,PhysRevLett.131.126502} and cold atom setups~\cite{PhysRevLett.94.040404,PhysRevA.78.010101}, and experimental signatures consistent with a LT have been recently reported in Josephson junction arrays~\cite{kuzmin2025observation}.
In one of the early mesoscopic proposals  Le Hur predicted a LT in a charge Kondo circuit induced by a fluctuating gate voltage governed by a resistor $R$ and acting as an effective dissipative environment \cite{PhysRevLett.92.196804}, at which a charge step in a quantum dot becomes infinitely sharp at zero temperature~\cite{PhysRevB.66.075318,PhysRevLett.95.086406}. Motivated by recent advances in charge Kondo circuits~\cite{article,PhysRevX.8.031075,
doi:10.1126/science.aan5592,Piquard_2023}, we discuss an explicit and  realistic realization of the  LT in a modified circuit.

The charge Kondo effect has been  envisioned~\cite{PhysRevB.48.11156,PhysRevB.51.1743,PhysRevLett.75.709} to arise from a twofold degeneracy in the charge states of a metallic island in the Coulomb blockade regime. Dynamics is induced by tunneling $t$ between the island and one or more leads. At weak tunneling the Kondo Hamiltonian takes the standard form $
H_K = \sum_{k\sigma}\epsilon_k\, c^\dagger_{k\sigma} c_{k\sigma}
+ J_\perp \left( S^+ s^-_0 + S^- s^+_0 \right)
+ J_z\, S^z s^z_0$, and is realized with impurity spin-1/2 $S^\alpha$ representing the two charge states. The (dimensionless) transverse and longitudinal exchange couplings $J_\perp, J_z$ have bare values $J_\perp \propto t$  and $J_z=0$. Here
$s^\alpha_0 $ is the spin density of the conduction electrons at the impurity site where the two spin species refer to states inside or outside the island, and we consider a single channel here.  The Kondo effect  is then captured by the RG equation
\be
\label{eq:RG0}
\frac{dJ_\perp}{dl} = J_\perp J_z , \qquad
\frac{dJ_z}{dl} = J_\perp^2 ,
\ee
where $l = \ln(D/D(\ell))$ is the RG flow parameter with $D$ the bare cutoff and $D(\ell)$ the running scale, predicting that $t$ flows to strong coupling with asymptotic isotropy $J_z = J_\perp$. These predictions were materialized in a sequence of experiments based on semiconductor nanostructures operating in the integer quantum Hall regime, where the  one- two- and three-channel charge Kondo models have been unambiguously realized~\cite{article,
doi:10.1126/science.aan5592,Piquard_2023}. Such devices generally referred to as charge-Kondo circuits were then identified as tunable quantum simulators of other interacting models, such as the problem of an impurity in a Luttinger liquid~\cite{PhysRevX.8.031075}. The physics of many such metallic islands had only started to be explored with a pioneering experiment on two islands~\cite{pouse2023quantum} and theory predictions for many islands~\cite{stäbler2024giantheatfluxeffect,Karki_2022,roche2025breakdownwiedemannfranzlawinteracting,hurvitz2025metallicislandarraysynthetic}.

\begin{figure}
    \centering
    \includegraphics[width=.7\linewidth]{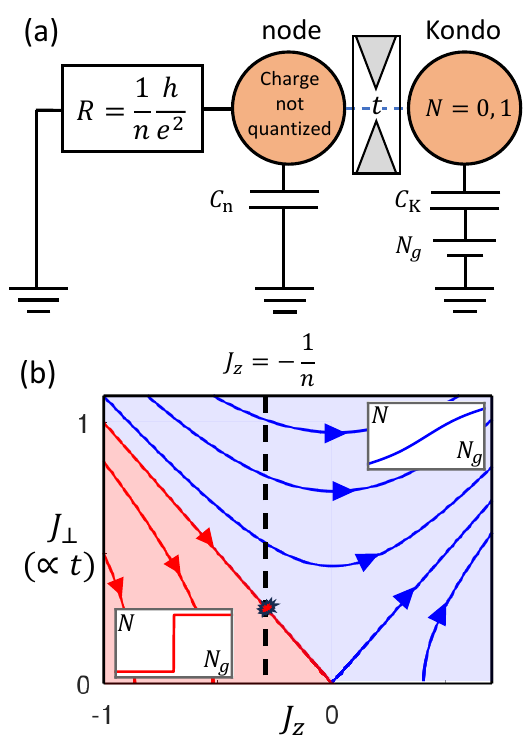}
    \caption{(a) Charge Kondo circuit with Luttinger-liquid interactions. A resistor is connected to the node (a large metallic dot without charge quantization), and electron tunneling occurs between this node and the Kondo metallic dot. (b) Phase diagram of the anisotropic Kondo model in Eq.~\eqref{eq:RG0}. The red (blue) region corresponds to the localized (delocalized) phase.}
    \label{fig:1}
\end{figure}

While the charge Kondo model in metallic islands has been widely studied~\cite{PhysRevB.52.16676,PhysRevLett.86.280,PhysRevLett.116.157202,PhysRevLett.116.157202,PhysRevB.97.085403,PhysRevLett.120.186801,PhysRevLett.125.026801,PhysRevB.102.041111}, the role of electron-electron interactions in the QPCs has only recently been explored~\cite{Nguyen_2020,PhysRevB.105.L121405,PhysRevB.107.L201402,PhysRevResearch.5.023019,paris2025threechannelchargekondomodel}. In this work we propose a modified charge Kondo circuit as a realization of LL interactions, see Fig.~\ref{fig:1}(a). In our circuit the Kondo island is coupled to another island denoted ``node" having sizable charging energy but yet no charge quantization by virtue of being connected to $n$-open channels. These $n$ open channels result in a quantized resistance to ground $R= \frac{h}{e^2} \frac{1}{n}$ which creates voltage fluctuations in the node and controls the effective LL parameter $K=\frac{1}{1+\frac{e^2}{h}R}=\frac{n}{n+1}$. We emphasize that the charging energy in the node is essential in our proposal making it distinct from Refs.~\cite{safi2004one,anthore2018circuit}: without charging energy the voltage fluctuations across the resistor would not affect the floating Kondo island. 
In fact, the capacitance of the node $C_n$ introduces an energy scale $\frac{\hbar}{R C_n}$ which acts as a high energy cutoff for the LL physics. Below this scale, the  RG equations become
\bea
\label{eq:RG}
\frac{dJ_\perp}{d l}=J_{\perp} J_z - \frac{1}{n} J_\perp , \qquad \frac{dJ_z}{d l}=J_\perp
^2,~~(n \ge 1),
\eea
with the same initial values. This is equivalent to Eq.~(\ref{eq:RG0}) with a negative initial value $J_z=-1/n$. As indicated in Fig.~\ref{fig:1}(b), now there is a quantum phase transition versus $t$: for bare values of $J_\perp$ below a critical value, the tunneling flows to zero. This corresponds to the LT, previously inaccessible in charge Kondo circuits. We will explicitly show how this transition can be detected via the charge susceptibility and entropy.

\begin{figure}
    \centering
    \includegraphics[width=\linewidth]{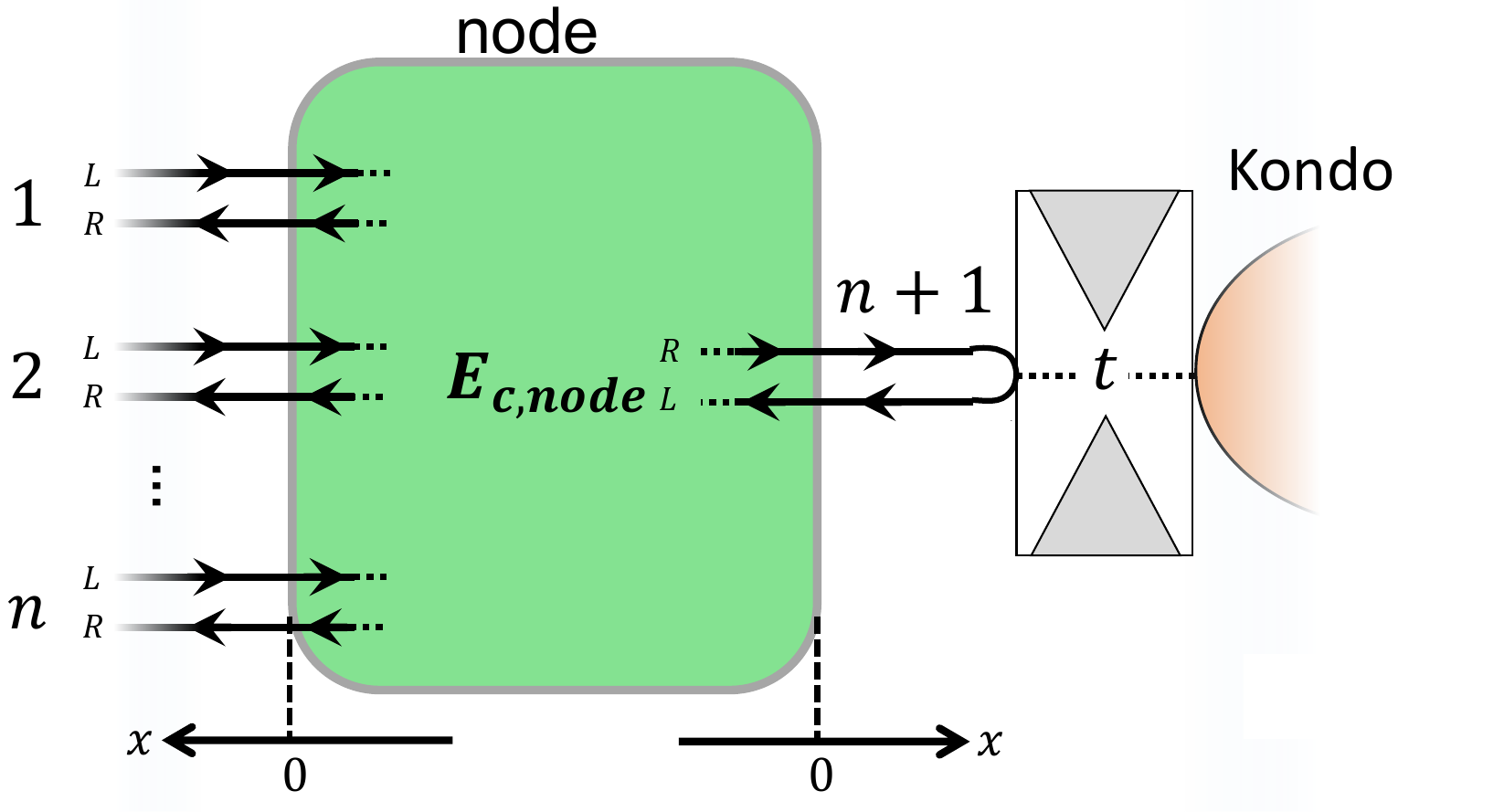}
    \caption{Circuit node as a metallic island with dominant charging energy, serving as a simulator of LL physics~\cite{lee2020fractional,PhysRevB.105.075433}. It connects to $n+1$ integer quantum Hall edge channels, one of which is partitioned by a QPC and coupled to a charge-Kondo island via a weak transmission amplitude $t$. 
    }
    \label{fig:2}
\end{figure}

 \textbf{Interacting circuit node as a simulator of LL physics:}
As shown in Fig.~\ref{fig:2}, the node in our circuit can be described by an island with $n+1$ integer edge channels (by channel we refer to a pair of counter-propagating edge modes), including $n$ open channels that realize the resistor $R=\frac{1}{n}\frac{h}{e^2}$, and one additional channel connected to the Kondo island via a weakly transmitting quantum point contact (QPC).

Before  taking into account reflection at the QPC, we consider the $n+1$'s channel as another open channel, and follow Refs.~\cite{lee2020fractional,PhysRevB.88.165307, PhysRevB.105.075433} to write the node Hamiltonian as $H_{node}=H_0+E_{c,node} \hat{N}_{node}^2$ where $E_{c,node}=\frac{e^2}{2C_{n}}$ is its charging energy and
\be
H_0=\sum_{j=1}^{n+1} \frac{\hbar v_F}{4\pi} \int_{-\infty}^\infty dx [(\partial_x \phi_{L,j})^2+(\partial_x \phi_{R,j})^2],
\ee
with 
\be
\hat{N}_{node}=\sum_{j=1}^{n+1} \int_{-\infty}^0 (\rho_{L,j} + \rho_{R,j})= \frac{1}{2\pi} \sum_{j=1}^{n+1} (\phi_{R,j}(0)-\phi_{L,j}(0)),
\ee
$\rho_{R/L,j}= \pm \frac{1}{2\pi} \partial_x \phi_{R/L,j}$, $[\phi_{\mu,j}(x),\phi_{\mu',j'}(x')]=\delta_{\mu \mu'} \delta_{jj'} \mu i \pi {\rm{sgn}}(x-x')$ ($\mu,\mu'=R/L=\pm$), and $\psi_{R/L}(x)= a^{-1/2} e^{i \phi_{R/L} (x)}$ is the electron operator with $a=\frac{2\pi \hbar v_F}{D}$.

We proceed as in Ref.~\cite{PhysRevB.105.075433} with a boundary condition relating the bosonic fields before and after the point $x=0$ where they interact through the charging energy. In particular, following the formalism of Ref.~\cite{hurvitz2025metallicislandarraysynthetic}, we write a boundary condition expressing the $2n+2$ outgoing fields in terms of  the $2n+2$ incoming fields,
\be
\label{eq:S}
\begin{pmatrix}
\Vec{\phi}_L(0^-) \\
\Vec{\phi}_R(0^+)
\end{pmatrix} =\hat{S} \begin{pmatrix}
\Vec{\phi}_L(0^+) \\
\Vec{\phi}_R(0^-)
\end{pmatrix}, ~~~\hat{S}=\begin{pmatrix}
1- \hat{P} & \hat{P} \\
\hat{P} & 1-\hat{P} 
\end{pmatrix},
\ee
where $\Vec{\phi}_{L/R}=(\phi_{L/R,1},\phi_{L/R,2},\dots \phi_{L/R,n+1})$ and $\hat{P}_{ij}=\frac{1}{n+1}$ [see Ref.~\cite{hurvitz2025metallicislandarraysynthetic} for $n=1$]. In the RHS of Eq.~(\ref{eq:S}) we have $n+1$ incoming fields from outside the node $\Vec{\phi}_L(0^+)$ as well as $n+1$ fields incoming  from inside the node $\Vec{\phi}_R(0^-)$. Similarly in the LHS we have $n+1$ fields outgoing into the node $\Vec{\phi}_L(0^-)$ as well as
$n+1$ outgoing fields towards the outside $\Vec{\phi}_R(0^+)$. So $\hat{S}$ is a $(2n+2) \times (2n+2)$ scattering matrix.
Physically, any incoming combination of $n+1$ bosonic states can be decomposed into a charge mode (corresponding to the $n+1-$dimensional normalized vector $\vec{v}_c=\frac{1}{\sqrt{n+1}} (1,1,\dots 1)$) which is fully reflected and $n$ neutral modes which are fully transmitted. $\hat{P}_{ij}=(\vec{v}_{c})_i(\vec{v}_{c})_j$ is the $(n+1) \times (n+1)$ projector into the space spanned by the charge vector. 

\begin{figure*}
    \centering
    \includegraphics[width=\textwidth]{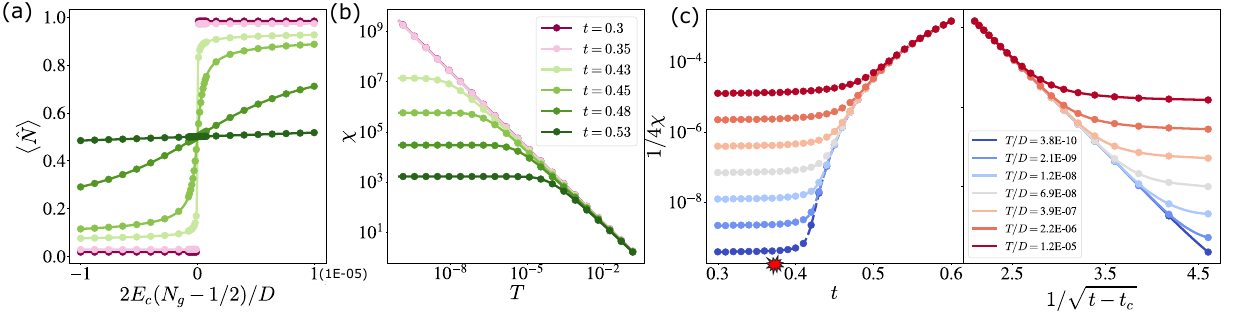}
    \caption{NRG results for $n=5$.  (a)$\langle\hat{N}\rangle$ versus $N_g$ for decreasing tunnelings. For $t>t_c$ (dark to light green), the charge state is delocalized, and we find a smooth crossover.
    For $t<t_c$ (dark to light pink), the charge state is localized, and we find a discontinuity at the transition point $N_g=1/2$.
    (b) Charge-susceptibility versus temperature for decreasing tunnelings. In the delocalized phase, $\chi$ saturates in the low temperature regime, while in the localized phase, $\chi$ keeps increasing down to $T=0$.
    (c) Left: inverse susceptibility versus $t$ for decreasing temperatures, for lines with colors changing from red to blue. Right: inverse susceptibility versus $1/\sqrt{t-t_c}$, where $t_c=0.3782$, for the same temperatures. In the limit $T\to 0$, the curves exhibit the KT scaling behavior $\chi^{-1}\sim \exp(-\mathrm{const}/\sqrt{t-t_c})$.}
    \label{fig:enter-label}
\end{figure*}

Now consider the QPC in the limit $t \to 0$ leading to full reflection at the point $x_{QPC}>0$ such that $\frac{2\pi \hbar v_F}{x_{QPC}} \gg k T$. In the analysis below, we take $T\to0$ and then $x_{QPC}\to 0$. This yields the identification
\be
\label{eq:identification}
\phi_{R,n+1}(0^+)=\phi_{L,n+1}(0^+) \equiv \phi.
\ee
Let us express $\phi$ in terms of only the incoming fields $\vec{I} = (\Vec{\phi}_L(0^+) ,
\Vec{\phi}_R(0^-) )$. Eq.~(\ref{eq:S}) gives
\bea
(n+1) \phi_{R,n+1}(0^+)&=&\sum_{j=1}^n \phi_{L,j}(0^+) + \phi_{L,n+1}(0^+) \nonumber \\
&-& \sum_{j=1}^n \phi_{R,j}(0^-)+n \phi_{R,n+1}(0^-).
\eea
Using Eq.~(\ref{eq:identification}) this gives 
\be
\label{eq:phi_d}
\phi= \vec{d} \cdot \vec{I},~~\vec{d}=\frac{1}{n}((1,1,\dots, 1,0),(n,-1,-1,\dots,-1)).
\ee

To study the effect of the interacting node on the charge-Kondo physics we only need to deduce the properties of the electron operator $\Psi_{node}=\psi_{R,n+1}(0^+) =\psi_{L,n+1}(0^+) =a^{-1/2} e^{i \phi}$ in the tunneling operator. Using the result Eq.~(\ref{eq:phi_d}), we see that $\phi$ is expressed in terms of a linear combination of independent incoming fields. We define a normalized right-moving field 
\be
\tilde{\phi}(x) =(\frac{\vec{d}}{|\vec{d}|}  [ \Theta(-x) + \hat{S}^{-1} \Theta(x) ] )\cdot (\vec{\phi}_L(-x),\vec{\phi}_R(x)). 
\ee
In the incoming region $x<0$ it is defined such that at $x=0^-$ it coincides with the normalized $\phi$ field. Its definition in the outgoing region $x>0$ is such that $\tilde{\phi}(0^-)=\tilde{\phi}(0^+)$. It is normalized as usual as 
$[\tilde{\phi}(x),\tilde{\phi}(x')]= i \pi {\rm{sgn}}(x-x')$ and  its Hamiltonian is $H_0[\tilde{\phi}]=\frac{\hbar v_F}{4\pi} \int_{-\infty}^\infty dx (\partial_x \tilde{\phi})^2$. The electron operator 
is given by
\be
\Psi_{node} =a^{-1/2} e^{i \phi} =a^{-1/2} e^{i |\vec{d}| \tilde{\phi}(0)},
\ee
and therefore has scaling dimension $\frac{1}{2} |\vec{d}|^2 = \frac{n^2+2n}{2n^2}$.

Including the Kondo island, the total Hamiltonian is
\bea
\label{eq:Psi}
H_{tot}&=&\frac{\hbar v_F}{4\pi} \int_{-\infty}^\infty dx (\partial_x \tilde{\phi})^2+ t a^{-1/2}(e^{i |\vec{d}| \tilde{\phi}(0)} \Psi(0)+h.c.) \nonumber \\ 
&+&\hbar v_F \int_{-\infty}^\infty dx \Psi^\dagger(x) i \partial_x \Psi(x) + E_c ( \hat{N}-N_g)^2, 
\eea  
where $\hat{N} = \int_{-\infty}^\infty dx \Psi^\dagger(x) \Psi(x)$. The scaling dimension of the electron operator at the Kondo island $\Psi(x)$ is 1/2. The total scaling dimension of the tunneling operator is then $\frac{1}{2}+\frac{n^2+2n}{2n^2}=1+\frac{1}{n}$. The effect of the $n \ge 1$ open channels is to turn the tunneling $t$ from being marginal to irrelevant, as described by the RG equation Eq.~(\ref{eq:RG}).

 \textbf{NRG results:}
We  apply numerical renormalization group (NRG) methods to identify the LT and compute the thermodynamic properties. We build on a  standard fermionic solution of the charge Kondo model~\cite{PhysRevB.68.041311, PhysRevB.70.201306}. We denote the fermions in the Kondo island by ``spin down" and replace the interacting node by a lead whose fermions are denoted by ``spin up", and write the Hamiltonian in the standard way~\cite{PhysRevB.68.041311, PhysRevB.70.201306}
\bea
H_{NRG} &=& \sum_{\sigma =\uparrow,\downarrow} \sum_k \epsilon_{k}\, c^\dagger_{\sigma k} c_{\sigma k} +  t \sum_{k, k'} c^\dagger_{\downarrow k} c_{\uparrow k'}\, \hat{N}^+ + \mathrm{H.c.} \nonumber \\
&+& E_c(\hat{N}-N_g)^2 + H_{\text{env}},
\label{eq:H_tot_large_r}
\eea
where $\hat{N}^\pm$ is the raising/lowering operator for $\hat{N}$, $[\hat{N},\hat{N}^\pm]=\pm \hat{N}^\pm$ and $c_{\alpha k}$ are  fermionic annihilation operators. In our calculations, we restrict our attention to two charge states only. This corresponds to the limit of large $E_c \gg k T$, in which $\hat{N}$ is dominated by two charge states and behaves like a spin-1/2 operator $S^z = \hat{N}-1/2$, $S^+=\hat{N}^+$. In order to include the effect of the LL interactions generated by the interacting node in our circuit, we add the last term $H_{\text{env}}$, describing an additional environmental degree of freedom. In terms of an additional normalized boson field, this term is given by
\begin{equation}
\label{eq:NRG_H_env}
H_{\text{env}} =\hbar v_F \alpha\, \hat{N}\, \partial_x \phi_{e}(0) + \frac{\hbar v_F}{4\pi} \int dx\, (\partial_x \phi_{e})^2,
\end{equation}
with $\alpha^2=2/n$~\cite{SM}.
This fermionic model is well-suited to NRG treatment, following the methods developed in~\cite{PhysRevB.68.041311, PhysRevB.70.201306}. To handle the coupling to the environment described by $H_{\text{env}}$, we follow the fermionic approach of~\cite{PhysRevLett.131.126502, PhysRevB.111.115117}, and map the bosonic environment into an effective fermionic bath.

In all our NRG results the charge counter $\hat{N}$ is restricted to two states. In Fig.~\ref{fig:enter-label}, we present NRG results for $n= 5$. In panel (a) we plot the expectation value $\langle \hat{N}\rangle$ versus $N_g$. We identify two phases: in the localized phase, the charge curve exhibits a discontinuous step as $T \to 0$, while in the delocalized phase, there is only a smooth crossover. In panel (b) we show the temperature dependence of the charge susceptibility $\chi = \partial \langle \hat{N} \rangle / \partial N_g |_{N_g=1/2}$ for decreasing tunnelings and observe that, in the localized phase ($t<t_c$), the susceptibility follows a Curie-like $1/T$ behavior down to $T \to 0$, while in the delocalized phase, it saturates at low temperatures. In panel (c) we plot the inverse charge susceptibility for decreasing temperatures and observe that as $T\to 0$ it continuously vanishes as $t \to t_c^+$. The phase transition described by the RG Eq.~(\ref{eq:RG}) leading to the phase diagram in Fig.~\ref{fig:1} is of Kosterlitz-Thouless (KT) type. The right panel of (c) demonstrates the KT 
scaling relation $\chi^{-1}\sim\exp(-\mathrm{const}/\sqrt{t-t_c})$, which also allows us to extract $t_c$. 

\begin{figure}
    \centering
    \includegraphics[width=.7\linewidth]{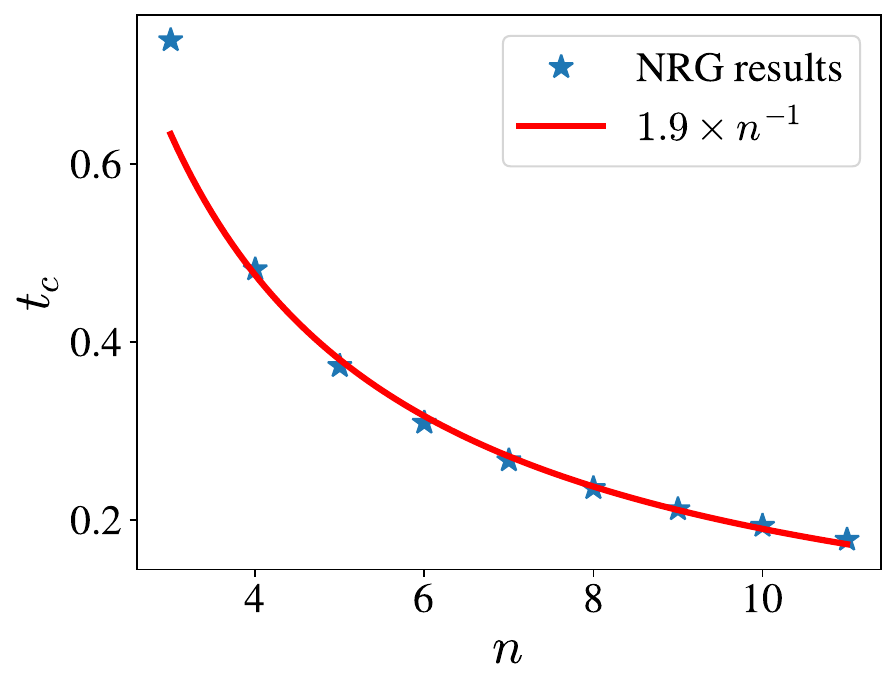}
    \caption{NRG results for the critical tunneling $t_c$ versus $n$, demonstrating  that $t_c \propto n^{-1}$ near the KT transition.}
    \label{fig:4}
\end{figure}

In Fig.~\ref{fig:4} we plot $t_c$ versus LL parameter for $n=3,4,\dots,11$. At large values of $n$ we approach the vicinity of the KT transition, and as indicated in Fig.~\ref{fig:1} the critical tunneling becomes linear in $1/n$, serving as another confirmation of the KT nature of the transition.

\begin{figure}
    \centering
    \includegraphics[width=.8\linewidth]{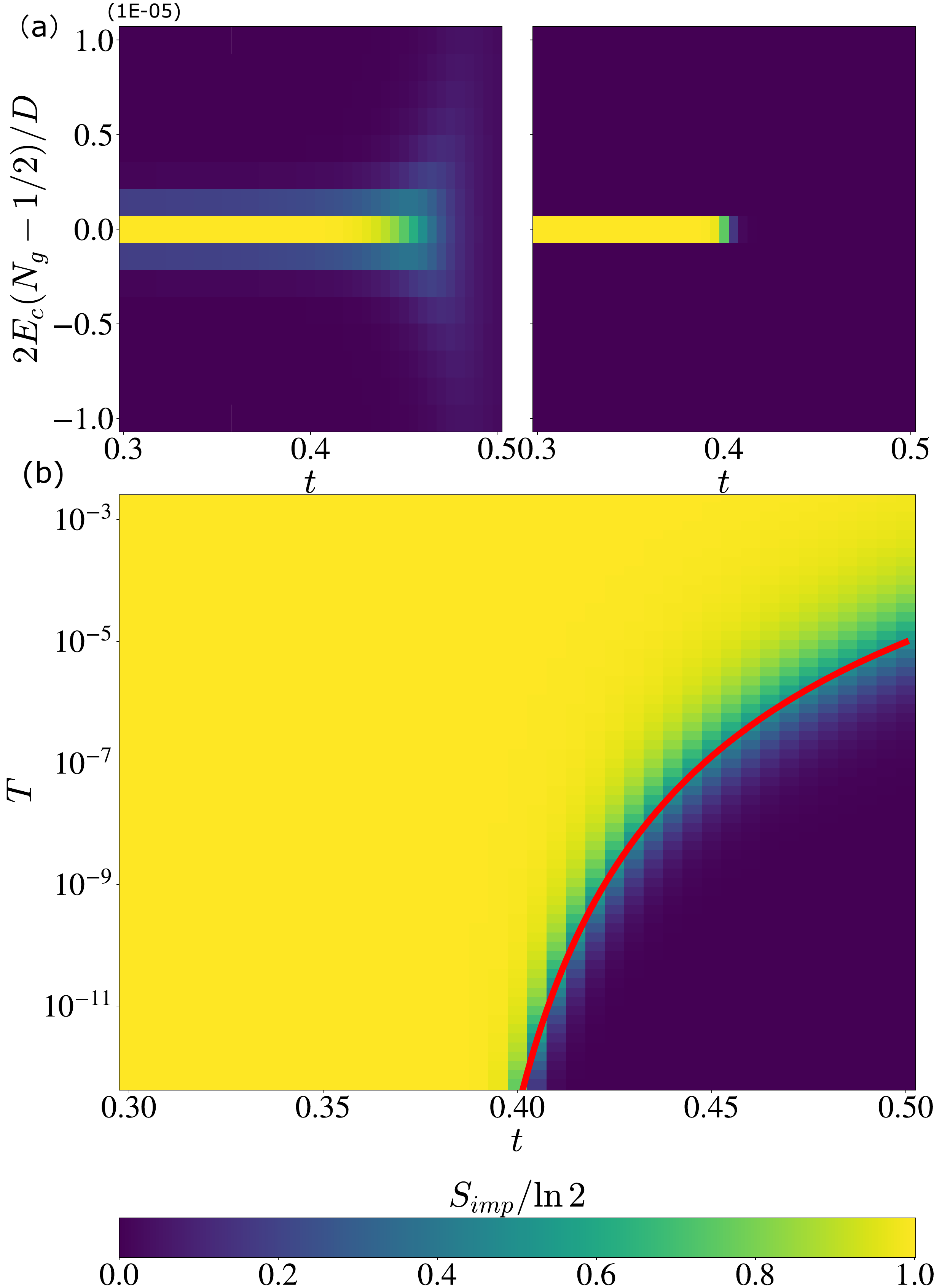}
    \caption{NRG results for the impurity entropy $S_{imp}$, obtained from the Hamiltonian Eq.~\ref{eq:H_tot_large_r} with $n=5$; see text for further discussion.}
    \label{fig:entropy}
\end{figure}

 \textbf{Thermodynamic entropy:}
In Fig.~\ref{fig:entropy} we show NRG results for the impurity entropy $S_{imp}$. 
In panel (a), we plot $S_{imp}$ versus $t$ and $N_g$ for $T=4\times 10^{-7}D$ (left) and $T=10^{-12}D$ (right). 
At the charge degeneracy point $N_g=1/2$, $S_{imp}$ shows a crossover from $\ln{2}$ to $0$ as $t$ increases. The crossover becomes sharper as $T$ lowers. Any deviation from the charge degeneracy point renders $S_{imp}$ zero in the low temperature regime.  
In panel (b), we plot $S_{imp}$ versus $t$ and $T$ at the charge degeneracy point. In the delocalized phase $t>t_c$, $S_{imp}$ decreases from $\ln{2}$ at high temperatures to $0$ at low temperatures. The crossover temperature $T_K$, which is proportional to the inverse susceptibility~\cite{PhysRevB.111.115117}, shows the same scaling behavior $T_K\sim \exp(-\mathrm{const}/\sqrt{t-t_c})$, as depicted by the red line. In the localized phase, $S_{imp}$ remains $\ln{2}$ down to zero temperature. 
The impurity entropy of the system can be measured with the Maxwell relation\cite{hartman2018direct,PhysRevLett.129.227702,child2022robust, PhysRevLett.123.147702,PhysRevLett.128.146803}. 

 \textbf{Discussion - Transport signatures:}
Imagine isolating a single pair of counter-propagating chiral modes in Fig.~(\ref{fig:2}) for a conductance measurement. For example, we can measure transport from mode $1L$ to mode $1R$ while the remaining $n-1$ channels are grounded. Is the DC conductance sensitive to the LT at $t=t_c$? The answer is immediately seen to be negative for the one-channel charge Kondo case that we consider, even without calculation: because the Kondo island is floating, no net current can flow into it, either in the localized phase $t \to 0$ or in the delocalized phase $t > t_c$.

Let us suggest two scenarios where a transport signature of the LT is expected without changing the number of Kondo channels. 

One scenario involves adding a weak link of the Kondo island to ground with $R_{weak~link} \gg h/e^2$. In this limit the model is still a one-channel charge Kondo model with a small perturbation. But this extra perturbation allows for a finite DC current flow into the Kondo island from one of the $n$ channels - only in the delocalized phase. 

Another possibility is to consider an interference experiment as in Ref.~\cite{Duprez_2019}. Imagine an interferometer between 
an external arm and an additional arm that enters our system through mode $1L$ and exits 
via mode $1R$. Does one expect interference as a function of $N_g$? Interference versus 
$N_g$ is expected only in the localized phase, and must disappear in the delocalized phase 
as $t$ is tuned below $t_c$. Yet, this effect is expected to hold for $n=1$, but is expected to vanish for $n >1$ ballistic channel because the outgoing electrons (in mode $1R$) lose their coherence versus mode $1L$~\cite{Sela_2023} irrespective of the LT.

 \textbf{Summary:}
Charge-Kondo circuits serve as highly tunable quantum simulators for nontrivial quantum-impurity problems and fractionalization phenomena. Their ability to emulate strongly interacting systems arises from the charging energy that governs the metallic islands forming the circuit nodes, as well as from the universal character of the links between these nodes, realized as quantum Hall  edge states. In this work, we broadened the scope of these quantum simulators to include the localization transition— a quantum phase transition that emerges in a wide variety of physical models, including the Kondo effect in the presence of Luttinger-liquid interactions. 

We note that a fully equivalent realization of the localization transition is provided by a QPC in the fractional quantum Hall regime. When the QPC separates a large fractional quantum Hall droplet from a finite region with non-negligible charging energy, the localization transition occurs as a function of the QPC’s reflection amplitude.

Finally, we note that while our model displays a phase transition at finite Kondo coupling, and hence reminds of the pseudo-gap Kondo model~\cite{withoff1990phase}, it is different. Particularly, the pseudogap Kondo model displays phases with fractional entropy~\cite{Gonzalez_Buxton_1998,Fritz_2004} while our model displays fixed point entropy of either $0$ or $\ln 2$.

Our results open several promising directions for future investigation. First, an experimental realization of our predictions appears to be well within reach. Second, many important cases remain unexplored, including localization transitions in multichannel Kondo models. Such generalizations—for example, a two-channel Kondo implementation of our circuit—would exhibit direct conductance signatures. Finally, as quantum-circuit platforms advance toward scalable architectures with many islands, the localization transitions identified here at the single-impurity level may give rise to new forms of extended, many-body quantum phase transitions.


 \textbf{Acknowledgments:}
ZM FP and ES gratefully acknowledge support from the European Research Council (ERC) under the European Union Horizon 2020 research and innovation programme under grant agreement No. 951541.


\begin{appendix}
\section{Derivation of $H_{NRG}$}
To map Eq.~(\ref{eq:Psi}) to Eqs.(\ref{eq:H_tot_large_r}) and (\ref{eq:NRG_H_env}) we perform the unitary transformation $H'=U^\dagger H U$ with $U=e^{i \beta \phi_e \hat{N}}$. Under this transformation
\bea
U^\dagger \partial_x \phi_e(x) U &=& \partial_x \phi_e(x) - 2 \pi \beta \hat{N} \delta(x)~\Rightarrow \nonumber \\
U^\dagger H_{\text{env}} \phi_e U &=& H_{\text{env}}- \hbar v_F \beta \hat{N} \partial_x \phi_e(0).
\eea
By selecting $\beta = \alpha$ the coupling term $\propto \alpha$ disappears from Eq.~(\ref{eq:NRG_H_env}) and reappears in the tunneling term with
\be
U^\dagger N^+  U = N^+ e^{- i \beta \phi_e(0)}.
\ee
The scaling dimension of the resulting extra vertex operator is $\Delta[e^{- i \beta \phi_e}] |_{\beta= \alpha}= \frac{\alpha^2}{2}$. With $\alpha^2=2/n$, the total scaling dimension of the tunneling terms in Eqs.~(\ref{eq:Psi}) and (\ref{eq:H_tot_large_r}) coincide with $1+1/n$.

\section{
Derivation of the anisotropic Kondo model from Eq.~\ref{eq:H_tot_large_r}}

We bosonize the fermion in the box ($\sigma=\downarrow$) and the fermion in the lead ($\sigma=\uparrow$) and combine the tunneling operator into
\be 
t e^{i (\phi_{\uparrow}(0)-\phi_{\downarrow}(0))} \hat{N}^+ e^{-i \alpha \phi_e(0)} =t e^{i \sqrt{2+\alpha^2} \phi_s(0)} \hat{N}^+ .
\ee
In the last equality we defined  a normalized chiral right moving spin field $\phi_s(x) = \frac{\phi_{\uparrow}(x)-\phi_{\downarrow}(x)-\alpha \phi_e(0)}{\sqrt{2+\alpha^2}}$.
Our starting point is the Hamiltonian
\bea
H_0 &=& \frac{\hbar v_F}{4 \pi} \int dx [(\partial_x \phi_\uparrow)^2+(\partial_x \phi_\downarrow)^2+(\partial_x \phi_e)^2] \nonumber \\
&+& t a^{-1/2}e^{\sqrt{2+\alpha^2} \phi_s(0)}\hat{N}^++ h.c..
\eea
We make a rotation of the 3 boson fields to a new basis with $\phi_s(x)$ being one of the basis elements. The scaling dimension of the tunneling operator is $\Delta[e^{\sqrt{2+\alpha^2} \phi_s(0)}]= \frac{1}{2} (2+\alpha^2)=1+ \frac{\alpha^2}{2}$ as expected. We now intend to change this scaling dimension to unity, as in the spin-flip term in the Kondo model. We do so using another unitary transformation
\be
U'=e^{- \gamma \phi_s(0) \hat{N}},
\ee
which takes
\be
{U'}^\dagger e^{\sqrt{2+\alpha^2} \phi_s(0)} \hat{N}^+  U'=e^{(\sqrt{2+\alpha^2} -\gamma )\phi_s(0)} \hat{N}^+.
\ee
Setting the scaling dimension to unity  - which is the scaling dimension of the spin-flip term in the Kondo Hamiltonian - is achieved by setting $\sqrt{2}=\sqrt{2+\alpha^2} -\gamma $. 
This generates from the kinetic energy a $J_z$ term in the Kondo Hamiltonian
\be
{U'}^\dagger H_0[\phi_s]  U'=H_0[\phi_s] + \hbar v_F \partial_x \phi_s(0) \gamma \hat{N}.
\ee
We thus obtain the anisotropic Kondo model, where the components of the spin densities of the conduction electrons are
\be
s^\pm (x) \sim e^{\mp i \sqrt{2} \phi_s(x)},~~~s^z(x) \sim \partial_x \phi_s(x).
\ee
which satisfy $SU(2)_1$ algebra. 
Therefore $J_z \propto \gamma = \sqrt{2+\alpha^2}-\sqrt{2}=\sqrt{2} \left( \sqrt{1+\frac{1}{n}} -1\right) = \frac{1}{\sqrt{2} n} + \mathcal{O}(\frac{1}{n^2})$.

We note that the unitary transformation performed using bosonization for small $n$ generates large coupling of the order of the band-width, therefore one can only rely on this result for $J_z$ to leading order in $1/n$.

\end{appendix}

\bibliography{bibliography}
\end{document}